\documentclass[eqsecnum,aps,twoside]{revtex4-2}

\usepackage{hyperref,amsfonts,amsmath,amssymb,bm}

\begin{document}

\author{Maxim Dvornikov}
\email{maxim.dvornikov@gmail.com}

\title{Quantization of massive fermions in vacuum and external fields}

\affiliation{Pushkov Institute of Terrestrial Magnetism, Ionosphere
and Radiowave Propagation (IZMIRAN),
108840 Moscow, Troitsk, Russia}

\begin{abstract}
We study massive Majorana neutrinos in background matter. Representing these particles in terms of Weyl spinors, we carry out their quantization. The propagators of these fields are also constructed. Then, we apply the Hamilton dynamics based formalism to describe massive Majorana neutrinos in matter on the classical level. Finally, we study a classical Dirac particle in vacuum, described with $c$-number variables, within the Hamiton formalism. Such a Dirac field is also canonically quantized.
\end{abstract}

\maketitle


\section{Introduction}

Neutrinos provide a unique tool to explore physics beyond the standard model. Despite the absolute scale of neutrino masses is unknown, we believe that neutrino masses are nonzero. Moreover, the question whether neutrinos are Dirac or Majorana particles is still open.

In the present work, we analyze some issues of the quantization of Majorana neutrinos in external fields. For the first time, the quantum treatment of massive Majorana neutrinos, represented in terms of Weyl spinors, was given in Ref.~\cite{Cas57}. Then, this formalism was used in Ref.~\cite{SchVal81} to study the interaction of Majorana neutrinos with magnetic fields. An alternative approach for the description of massive Majorana neutrinos in vacuum and various external fields, which is based on the Hamilton dynamics, was developed in Refs.~\cite{Dvo12a,Dvo12b}. The description of classical Majorana neutrinos in external fields, involving anticommuting Grassmann variables, was performed in Ref.~\cite{DvoGit13}. The canonical quantization of these particles was also made in Ref.~\cite{DvoGit13}.

We carry out a detailed study of Majorana neutrinos in background matter in Sec.~\ref{sec:QUANT} including the quantization of such particles. The propagators of Majorana neutrinos are derived in Sec.~\ref{sec:PROP}. This research is motivated by the recent result in Ref.~\cite{Dvo25} that the quantum field theory description of neutrino flavor oscillations in matter is likely to be applicable for Majorana particles. However, the quantization and the propagators derivation in Ref.~\cite{Dvo25} was sketchy. In the present work, we fill in this gap.

The rest of this work, namely Secs.~\ref{sec:CLMAJ}-\ref{sec:QUANTDIR}, is devoted to the development of the formalism proposed in Refs.~\cite{Dvo12a,Dvo12b}. In Sec.~\ref{sec:CLMAJ}, we clarify some issues in the classical field theory description of these Majorana neutrinos in matter. Then, in Sec.~\ref{sec:DIRCLASS}, we apply the proposed method to study a classical Dirac fermion in vacuum. Finally, in Sec.~\ref{sec:QUANTDIR}, we quantize the Dirac field which is constructed in Sec.~\ref{sec:DIRCLASS}. Eventually, we conclude in Sec.~\ref{sec:CONCL}.

\section{Quantization of massive Majorana neutrinos in matter\label{sec:QUANT}}

Let us consider a single massive Majorana neutrino represented in
terms of the two component Weyl spinor $\eta$. If such a neutrino
interacts with background matter, the Lagrangian for this system
has the form,
\begin{equation}\label{eq:WeylLagr}
  \mathcal{L}=\mathrm{i}\eta^{\dagger}(\sigma^{\mu}\partial_{\mu})\eta-g\eta^{\dagger}\eta
  +\frac{\mathrm{i}}{2}m\eta^{\dagger}\sigma_{2}\eta^{*}-\frac{\mathrm{i}}{2}m\eta^{\mathrm{T}}\sigma_{2}\eta,
\end{equation}
where $\sigma^{\mu}=(1,-\bm{\sigma})$ is the relativistic generalization
of the Pauli matrices $\bm{\sigma}=(\sigma_{1},\sigma_{2},\sigma_{3})$
and $m$ is the mass of $\eta$. The explicit form of the effective
potentials of the neutrino interaction with background matter, $g$,
for different types of neutrinos are given in Ref.~\cite{Dvornikov:2002rs}. Based on Eq.~(\ref{eq:WeylLagr}),
we obtain the wave equation for $\eta$
\begin{equation}\label{eq:weqWeyl}
  \mathrm{i}\partial_{t}\eta+[(\bm{\sigma}\mathbf{p})-g]\eta+\mathrm{i}m\sigma_{2}\eta^{*}=0,
\end{equation}
where $\mathbf{p}=-\mathrm{i}\nabla$ is the momentum operator.

The exact solution of Eq.~(\ref{eq:weqWeyl}) has the form,
\begin{align}\label{eq:Weylsol}
  \eta(\mathbf{x},t)= & \int\frac{\mathrm{d}^{3}p}{(2\pi)^{3/2}}
  \Big[
    \lambda_{-}
    \left(
      a_{-}(\mathbf{p})w_{-}e^{-\mathrm{i}E_{-}t+i\mathbf{px}}+A_{-}a_{-}^{\dagger}(\mathbf{p})w_{+}e^{\mathrm{i}E_{-}t-\mathrm{i}\mathbf{px}}
    \right)
    \nonumber
    \\
    & +
    \lambda_{+}
    \left(
      A_{+}a_{+}(\mathbf{p})w_{+}e^{-\mathrm{i}E_{+}t+\mathrm{i}\mathbf{px}}+a_{+}^{\dagger}(\mathbf{p})w_{-}e^{\mathrm{i}E_{+}t-\mathrm{i}\mathbf{px}}  
    \right)
  \Big],
\end{align}
where 
\begin{equation}
  E_{\pm}=\sqrt{m^{2}+(p\mp g)^{2}},
\end{equation}
are the energies of right and left polarized neutrinos, $a_{\pm}^{\dagger}(\mathbf{p})$
and $a_{\pm}(\mathbf{p})$ are the creation and annihilation operators
corresponding to the helicities $\pm$ and the momentum $\mathbf{p}$.
The coefficients $\lambda_{\pm}$ and $A_{\pm}$ in Eq.~(\ref{eq:Weylsol})
are
\begin{equation}\label{eq:lamA}
  \lambda_{\pm}^{2}=\frac{E_{\pm}+p\mp g}{2E_{\pm}},\quad A_{\pm}=\mp\frac{m}{E_{\pm}+p\mp g}.
\end{equation}
The helicity amplitudes
\begin{equation}\label{eq:helamp}
  w_{+}=
  \left(
    \begin{array}{c}
      e^{-\mathrm{i}\phi/2}\cos\vartheta/2 \\
      e^{\mathrm{i}\phi/2}\sin\vartheta/2
    \end{array}
  \right),
  \quad
  w_{-}=
  \left(
    \begin{array}{c}
      -e^{-\mathrm{i}\phi/2}\sin\vartheta/2 \\
      e^{\mathrm{i}\phi/2}\cos\vartheta/2
    \end{array}
  \right),
\end{equation}
obey the relations
\begin{equation}\label{eq:helampprop}
  w_{\pm}(-\mathbf{p})=\mathrm{i}w_{\mp}(\mathbf{p}),
  \quad
  w_{\pm}\otimes w_{\pm}^{\dagger}=\frac{1}{2}\left[1\pm\bm{\sigma}\hat{p}\right],
  \quad
  w_{\pm}\otimes w_{\mp}^{\mathrm{T}}=\pm\frac{1}{2}\left(1\pm\bm{\sigma}\hat{p}\right)\mathrm{i}\sigma_{2},
\end{equation}
which can be checked by means direct calculations using Eq.~(\ref{eq:helamp}).
The angles $\vartheta$ and $\phi$ in Eq.~(\ref{eq:helamp}) fix
the direction of the neutrino momentum $\mathbf{p}$.

Based on Eq.~(\ref{eq:WeylLagr}), we get that the canonical momentum
is
\begin{equation}
  \pi(\mathbf{x},t)=\frac{\delta\mathcal{L}}{\delta\dot{\eta}(\mathbf{x},t)}=\mathrm{i}\eta^{*}(\mathbf{x},t).
\end{equation}
Since the neutrinos are fermions, the quantum Poisson bracket is the
equal time anticommutator $\left\{ \eta(\mathbf{x},t),\pi(\mathbf{y},t)\right\} =\mathrm{i}\delta(\mathbf{x}-\mathbf{y})$
or $\left\{ \eta(\mathbf{x},t),\eta^{*}(\mathbf{y},t)\right\} =\delta(\mathbf{x}-\mathbf{y})$.
Using Eq.~(\ref{eq:Weylsol}), one gets that this relation holds
true if
\begin{equation}\label{eq:aadeg}
  \left\{ a_{\pm}(\mathbf{p}),a_{\pm}^{\dagger}(\mathbf{q})\right\} =\delta(\mathbf{p}-\mathbf{q}),
\end{equation}
with other anticommutators being equal to zero.

The energy-momentum tensor of $\eta$ reads
\begin{equation}\label{eq:emt}
  T^{\mu\nu}=\frac{\partial\mathcal{L}}{\partial(\partial_{\mu}\eta)}\partial^{\nu}\eta-\eta^{\mu\nu}\mathcal{L}
  =\mathrm{i}\eta^{\dagger}\sigma^{\mu}\partial^{\nu}\eta-\eta^{\mu\nu}\mathcal{L}.
\end{equation}
Based on Eq.~(\ref{eq:emt}), one obtains the energy density in the
form,
\begin{equation}\label{eq:T00}
  T^{00}=\mathrm{i}\eta^{\dagger}(\bm{\sigma}\nabla)\eta+g\eta^{\dagger}\eta
  -\frac{\mathrm{i}}{2}m\eta^{\dagger}\sigma_{2}\eta^{*}+\frac{\mathrm{i}}{2}m\eta^{\mathrm{T}}\sigma_{2}\eta.
\end{equation}
Using Eqs.~(\ref{eq:weqWeyl}) and~(\ref{eq:T00}), as well as the
wave equation for $\eta^{*}$,
\begin{equation}
  \mathrm{i}\dot{\eta}^{*}+(\bm{\sigma}^{*}\mathbf{p})\eta^{*}+g\eta^{*}=\mathrm{i}m\sigma_{2}\eta,
\end{equation}
we derive the total energy of the Weyl field in the form,
\begin{align}\label{eq:E}
  \mathcal{E}= & \int\mathrm{d}^{3}xT^{00}=\int\mathrm{d}^{3}x
  \bigg[
    \mathrm{i}\eta^{\dagger}(\bm{\sigma}\nabla)\eta+g\eta^{\dagger}\eta+\frac{1}{2}\eta^{\mathrm{T}}
    \left(
      \mathrm{i}\dot{\eta}^{*}-\mathrm{i}(\bm{\sigma}^{*}\nabla)\eta^{*}+g\eta^{*}
    \right)
    \nonumber
    \\
    & +
    \frac{1}{2}\eta^{\mathrm{\dagger}}
    \left(
      \mathrm{i}\dot{\eta}-\mathrm{i}(\bm{\sigma}\nabla)\eta-g\eta^{*}
    \right)
  \bigg]
  =
  \frac{\mathrm{i}}{2}\int\mathrm{d}^{3}x\left(\eta^{\dagger}\dot{\eta}-\dot{\eta}^{\dagger}\eta\right).
\end{align}
Analogously, we get that $T^{0i}=\mathrm{i}\eta^{\dagger}\partial^{i}\eta$.
Hence, the total momentum of $\eta$, which is $P_{i}=\smallint\mathrm{d}^{3}xT^{0i}$,
reads
\begin{equation}\label{eq:P}
  \bm{P}=-\frac{\mathrm{i}}{2}\int\mathrm{d}^{3}x
  \left[
    \eta^{\dagger}\nabla\eta-(\nabla\eta^{\dagger})\eta
  \right].
\end{equation}
In Eqs.~(\ref{eq:E}) and~(\ref{eq:P}), we integrate by parts and
take into account that $\eta$ is the anticommuting field.

Finally, with help of Eqs.~(\ref{eq:lamA}) and (\ref{eq:helampprop}), we transform Eq.~(\ref{eq:E}) to
\begin{equation}\label{eq:Esq}
  \mathcal{E}=\int\mathrm{d}^{3}p\left[E_{-}a_{-}^{\dagger}(\mathbf{p})a_{-}(\mathbf{p})+E_{+}a_{+}^{\dagger}(\mathbf{p})a_{+}(\mathbf{p})\right]
  +\text{vacuum terms}.
\end{equation}
Analogously, Eq.~(\ref{eq:P}) can be rewritten in the form,
\begin{equation}\label{eq:Psq}
  \bm{P}=\int\mathrm{d}^{3}p\,\mathbf{p}\left[a_{-}^{\dagger}(\mathbf{p})a_{-}(\mathbf{p})+a_{+}^{\dagger}(\mathbf{p})a_{+}(\mathbf{p})\right]
  +\text{vacuum terms}.
\end{equation}
Equations~(\ref{eq:Esq}) and~(\ref{eq:Psq}), together with Eq.~(\ref{eq:aadeg}),
finalize the canonical quantization of Weyl neutrinos in background
matter.

\section{Propagators of massive Majorana neutrinos in matter\label{sec:PROP}}

We can define two types of propagators for Weyl neutrinos. They are
\begin{align}\label{eq:Sxy}
  S(x-y)= &
  \left\langle 0\left|T[\eta(x)\eta^{\dagger}(y)]\right|0\right\rangle
  =\theta(x_{0}-y_{0})\left\langle 0\left|\eta(x)\eta^{\dagger}(y)\right|0\right\rangle
  \nonumber
  \\
  & -
  \theta(y_{0}-x_{0})\left\langle 0\left|\eta^{*}(y)\eta^{\mathrm{T}}(x)\right|0\right\rangle,
\end{align}
and
\begin{align}\label{eq:tildeSxy}
  \tilde{S}(x-y)= &
  \left\langle 0\left|T[\eta(x)\eta^{\mathrm{T}}(y)]\right|0\right\rangle
  =\theta(x_{0}-y_{0})\left\langle 0\left|\eta(x)\eta^{\mathrm{T}}(y)\right|0\right\rangle
  \nonumber
  \\
  & -
  \theta(y_{0}-x_{0})\left\langle 0\left|\eta(y)\eta^{\mathrm{T}}(x)\right|0\right\rangle,
\end{align}
where $\theta(t)$ is the Heaviside step function.

First, we study $S(x)$ in Eq.~(\ref{eq:Sxy}). Using Eq.~(\ref{eq:Weylsol}),
we obtain the vacuum mean values as
\begin{align}\label{eq:0etaeta0}
  \left\langle 0\left|\eta(x)\eta^{\dagger}(y)\right|0\right\rangle  & =
  \int\frac{\mathrm{d}^{3}p}{(2\pi)^{3}}
  \Big[
    \lambda_{-}^{2}\left(w_{-}\otimes w_{-}^{\dagger}\right)e^{-\mathrm{i}E_{-}(x_{0}-y_{0})}
    \notag
    \\
    & +
    \lambda_{+}^{2}A_{+}^{2}\left(w_{+}\otimes w_{+}^{\dagger}\right)e^{-\mathrm{i}E_{+}(x_{0}-y_{0})}
  \Big]
  e^{\mathrm{i}\mathbf{p}(\mathbf{x}-\mathbf{y})},
  \nonumber
  \\
  \left\langle 0\left|\eta^{\dagger}(y)\eta(x)\right|0\right\rangle  & =
  \int\frac{\mathrm{d}^{3}p}{(2\pi)^{3}}
  \Big[
    \lambda_{+}^{2}\left(w_{-}\otimes w_{-}^{\dagger}\right)e^{\mathrm{i}E_{+}(x_{0}-y_{0})}
    \notag
    \\
    & +
    \lambda_{-}^{2}A_{-}^{2}\left(w_{+}\otimes w_{+}^{\dagger}\right)e^{\mathrm{i}E_{-}(x_{0}-y_{0})}
  \Big]
  e^{-\mathrm{i}\mathbf{p}(\mathbf{x}-\mathbf{y})}.
\end{align}
Based on Eqs.~(\ref{eq:helampprop}) and~(\ref{eq:0etaeta0}), we
rewrite $S(x)$ in the form,
\begin{align}\label{eq:Sfin}
  S(x)= & \int\frac{\mathrm{d}^{3}p}{(2\pi)^{3}}\frac{e^{\mathrm{i}\mathbf{px}}}{2p}
  \big\{
    [p-(\bm{\sigma}\mathbf{p})]\lambda_{-}^{2}\left(e^{-\mathrm{i}E_{-}t}\theta(t)-e^{\mathrm{i}E_{-}t}\theta(-t)A_{-}^{2}\right)
    \nonumber
    \\
    & -
    [p+(\bm{\sigma}\mathbf{p})]\lambda_{+}^{2}\left(e^{\mathrm{i}E_{+}t}\theta(-t)-e^{-\mathrm{i}E_{+}t}\theta(t)A_{+}^{2}\right)]
  \big\}
  \nonumber
  \\
  & =
  \frac{\mathrm{i}}{2}\int\frac{\mathrm{d}^{4}p}{(2\pi)^{4}}e^{-\mathrm{i}px}
  \bigg\{
    \bigg[
      \lambda_{-}^{2}\left(\frac{1}{p_{0}-E_{-}+\mathrm{i}0}+\frac{A_{-}^{2}}{p_{0}+E_{-}-\mathrm{i}0}\right)
      \nonumber 
      \\
      & +
      \lambda_{+}^{2}\left(\frac{1}{p_{0}+E_{+}-\mathrm{i}0}+\frac{A_{+}^{2}}{p_{0}-E_{+}+\mathrm{i}0}\right)
    \bigg]
    \nonumber
    \\
    & -
    (\bm{\sigma}\hat{p})
    \bigg[
      \lambda_{-}^{2}\left(\frac{1}{p_{0}-E_{-}+\mathrm{i}0}+\frac{A_{-}^{2}}{p_{0}+E_{-}-\mathrm{i}0}\right)
      \nonumber
      \\
      & -
      \lambda_{+}^{2}\left(\frac{1}{p_{0}+E_{+}-\mathrm{i}0}+\frac{A_{+}^{2}}{p_{0}-E_{+}+\mathrm{i}0}\right)
    \bigg]
  \bigg\},
\end{align}
where
\begin{equation}\label{eq:Heav}
  e^{\pm\mathrm{i}Et}\theta(\mp t)
  =\pm\frac{1}{2\pi\mathrm{i}}\int_{-\infty}^{+\infty}\frac{e^{-\mathrm{i}p_{0}t}\mathrm{d}p_{0}}{p_{0}\pm E\mp\mathrm{i}0},
\end{equation}
is the representation of the Heaviside function. In Eqs.\ (\ref{eq:Sfin})
and~(\ref{eq:Heav}), $\mathrm{i}0$ means a small imaginary term.

The Fourier image of $S$ in Eq.~(\ref{eq:Sfin}) for ultrarelativistic
neutrinos has the form,
\begin{equation}
  S_{\mathrm{L}}(p_{0},\mathbf{p})\approx\frac{1-(\bm{\sigma}\hat{p})}{2(p_{0}-E_{-}+\mathrm{i}0)}.
\end{equation}
It is this propagator which was used in Ref.~\cite{Dvo25} to study
neutrino flavor oscillations in background matter.

The propagator $\tilde{S}(x)$ in Eq.~(\ref{eq:tildeSxy}) can be
studied analogously. We just present the final result,
\begin{equation}\label{eq:tildeSfin}
  \tilde{S}(x)=\frac{m\sigma_{2}}{2}\int\frac{\mathrm{d}^{4}p}{(2\pi)^{4}}e^{-\mathrm{i}px}
  \left[
    \frac{\left(1-\bm{\sigma}\hat{p}\right)}{p_{0}^{2}-E_{-}^{2}+\mathrm{i}0}+\frac{\left(1+\bm{\sigma}\hat{p}\right)}{p_{0}^{2}-E_{+}^{2}+\mathrm{i}0}
  \right].
\end{equation}
If we consider Weyl neutrinos in vacuum, then $E_{-}=E_{+}=\sqrt{p^{2}+m^{2}}$.
In this case, the propagators in Eqs.~(\ref{eq:Sfin}) and~(\ref{eq:tildeSfin})
become
\begin{equation}\label{eq:Svac}
  S(x) =\int\frac{\mathrm{d}^{4}p}{(2\pi)^{4}}\frac{\mathrm{i}\tilde{\sigma}_{\mu}p^{\mu}e^{-\mathrm{i}px}}{p^{2}-m^{2}+\mathrm{i}0},
  \quad
  \tilde{S}(x) =\int\frac{\mathrm{d}^{4}p}{(2\pi)^{4}}\frac{m\sigma_{2}e^{-\mathrm{i}px}}{p^{2}-m^{2}+\mathrm{i}0},
\end{equation}
where $\tilde{\sigma}^{\mu}=(1,\bm{\sigma})$ and $p^2 = p_0^2 - \mathbf{p}^2$. The propagators in
Eq.~(\ref{eq:Svac}) coincide with the known results given, e.g.,
in Ref.~\cite{FukYan03}.

\section{Classical field theory description of Majorana neutrinos\label{sec:CLMAJ}}

In Sec.~\ref{sec:QUANT}, we assume that, before the quantization, the field $\eta$ is represented in terms of the anticommuting Grassmann variables. Otherwise the mass term in Eq.~\eqref{eq:WeylLagr} is zero identically for the commuting $c$-number wave functions $\eta$ and $\eta^*$. In fact, it was the reason for the claim in Ref.~\cite{SchVal81} that classical massive neutrinos do not exist.

We demonstrated in Refs.~\cite{Dvo12a,Dvo12b} that the classical field theory of Majorana neutrinos is still possible in frames of the Hamilton formalism. Indeed, let us consider the Hamiltonian $H_\mathrm{M}$ of the $c$-number commuting canonical variables $(\eta,\pi)$ and $(\eta^*,\pi^*)$,
\begin{align}\label{Hamclass}
  H_\mathrm{M} = & \int \mathrm{d}^3 x
  \Big\{
    \pi^\mathrm{T} (\bm{\sigma}\nabla) \eta -
    (\eta^{*{}})^\mathrm{T} (\bm{\sigma}\nabla) \pi^{*{}}
    + m
    \left[
      (\eta^{*{}})^\mathrm{T} \sigma_2 \pi +
      (\pi^{*{}})^\mathrm{T} \sigma_2 \eta
    \right]
    \notag
    \\
    & -
    \mathrm{i} g [\pi^\mathrm{T} \eta 
    - (\eta^{*{}})^\mathrm{T} \pi^{*{}}]
  \Big\},
\end{align}
where $g$ is the effective potential for the massive neutrino interaction with background matter, as in Sec.~\ref{sec:QUANT}.

The canonical equations for $(\eta,\pi)$, based on Eq.~\eqref{Hamclass}, read,
\begin{align}
  \label{etaclass}
  \dot{\eta}  = & \frac{\delta H_\mathrm{M}}{\delta \pi} =
  (\bm{\sigma}\nabla)\eta - m \sigma_2 \eta^{*{}}
  - \mathrm{i} g \eta,
  \\
  \label{piclass}
  \dot{\pi}  = & - \frac{\delta H_\mathrm{M}}{\delta \eta} =
  (\bm{\sigma}^{*{}}\nabla)\pi + m \sigma_2 \pi^{*{}}
  + \mathrm{i} g \pi,
\end{align}
and
\begin{align}
  \label{eta*class}
  \dot{\eta}^*  = & \frac{\delta H_\mathrm{M}}{\delta \pi^*} =
  (\bm{\sigma}^* \nabla)\eta^* + m \sigma_2 \eta
  + \mathrm{i} g \eta^*,
  \\
  \label{pi*class}
  \dot{\pi}^*  = & - \frac{\delta H_\mathrm{M}}{\delta \eta^*} =
  (\bm{\sigma}\nabla)\pi^* - m \sigma_2 \pi
  - \mathrm{i} g \pi^*,
\end{align}
are the canonical equations for $(\eta^*,\pi^*)$.

One can see that Eqs.~\eqref{etaclass} and~\eqref{eta*class} are identical to Eq.~\eqref{eq:weqWeyl}. If we take that $\pi = -\mathrm{i}\sigma_2 \xi$, we transform Eq.~\eqref{piclass} to the form,
\begin{equation}\label{eq:xiWeyl}
  \mathrm{i}\partial_{t}\xi-[(\bm{\sigma}\mathbf{p})-g]\xi-\mathrm{i}m\sigma_{2}\xi^{*}=0.
\end{equation}
Equation~\eqref{eq:xiWeyl} describes the evolution of the right-handed massive Majorana field $\xi$ in background matter, which corresponds to an antineutrino. It can be derived from the corresponding Dirac equation if we take that the bi-spinor is represented in the form~\cite{Dvo12b}, $\psi^\mathrm{T} = (\xi, -\mathrm{i}\sigma_2 \xi^*)$, which obeys the Majorana condition.

One can see that Eqs.~\eqref{etaclass} and~\eqref{eta*class} do not involve $\pi$ and $\pi^*$. It means that one cannot use them as the Legendre transformation to construct the Lagrangian. The same problem refers to Eqs.~\eqref{piclass} and~\eqref{pi*class}.

However, following Ref.~\cite{FadJac88}, we can construct the extended Lagrangian,
\begin{align}\label{extLagr}
  \mathcal{L}_\mathrm{R} = &
  \pi^\mathrm{T} \dot{\eta} + (\pi^{*{}})^\mathrm{T} \dot{\eta}^{*{}} -
  \left[
    \pi^\mathrm{T} (\bm{\sigma}\nabla) \eta -
    (\eta^{*{}})^\mathrm{T} (\bm{\sigma}\nabla) \pi^{*{}}
  \right]
  \notag
  \\
  & -
  m
  \left[
    (\eta^{*{}})^\mathrm{T} \sigma_2 \pi +
    (\pi^{*{}})^\mathrm{T} \sigma_2 \eta
  \right]
  + \mathrm{i} g [\pi^\mathrm{T} \eta - (\eta^{*{}})^\mathrm{T} \pi^{*{}}],
\end{align}
which can be formally obtained on the basis of Eq.~\eqref{Hamclass}: $\mathcal{L}_\mathrm{R} = \pi^\mathrm{T} \dot{\eta} + (\pi^{*{}})^\mathrm{T} \dot{\eta}^{*{}} - \mathcal{H}$. We treat $\eta$, $\eta^*$, $\pi$, and $\pi^*$ as independent variables in Eq.~\eqref{extLagr}.

Using the Euler-Lagrange equation for $\eta$,
\begin{equation}\label{ELe}
  \frac{\partial}{\partial t}
  \frac{\partial \mathcal{L}_\mathrm{R}}{\partial \dot{\eta}} +
  \nabla
  \frac{\partial \mathcal{L}_\mathrm{R}}{\partial \nabla \eta} =
  \frac{\partial \mathcal{L}_\mathrm{R}}{\partial \eta},
\end{equation}
we reproduce Eq.~\eqref{piclass}. 
Analogously, we can reproduce Eq.~\eqref{pi*class} by considering Eq.~\eqref{ELe} with respect to $\eta^*$.
Thus, one can treat $\mathcal{L}_\mathrm{R}$ as the Lagrangian for right-handed Majorana antineutrinos.

Analogously, we can define the additional Lagrangian
\begin{align}\label{extLagrL}
  \mathcal{L}_\mathrm{L} = &
  \eta^\mathrm{T} \dot{\pi} + (\eta^{*{}})^\mathrm{T} \dot{\pi}^{*{}} -
  \left[
    \eta^\mathrm{T} (\bm{\sigma}^\mathrm{T}\nabla) \pi -
    (\pi^{*{}})^\mathrm{T} (\bm{\sigma}^\mathrm{T}\nabla) \eta^{*{}}
  \right]
  \notag
  \\
  & +
  m
  \left[
    (\eta^{*{}})^\mathrm{T} \sigma_2 \pi +
    (\pi^{*{}})^\mathrm{T} \sigma_2 \eta
  \right]
  - \mathrm{i} g [\pi^\mathrm{T} \eta - (\eta^{*{}})^\mathrm{T} \pi^{*{}}].
\end{align}
Applying Eq.~\eqref{ELe} for $\mathcal{L}_\mathrm{L}$ with respect to $\pi$ and $\pi^*$, we reproduce Eqs.~\eqref{etaclass} and~\eqref{eta*class}, respectively. Thus, Eq.~\eqref{extLagrL} can be regarded as the Lagrangian for left-handed Majorana neutrinos.

We notice that $\mathcal{L}_\mathrm{R}$ and $\mathcal{L}_\mathrm{L}$ in Eqs.~\eqref{extLagr} and~\eqref{extLagrL} do not contain the factor $1/2$ which was improperly written down in Ref.~\cite[Eq.~(14)]{Dvo12a}. Finally, we mention that considering both $\mathcal{L}_\mathrm{R}$ and $\mathcal{L}_\mathrm{L}$ in Eqs.~\eqref{extLagr} and~\eqref{extLagrL} is analogous to the Routh formalism in classical mechanics (see, e.g., Ref.~\cite{LanLif76}).

\section{Classical field theory of a Dirac fermion\label{sec:DIRCLASS}}

It is interesting to mention that the formalism developed in Sec.~\ref{sec:CLMAJ} can be applied for the studies of Dirac femions which correspond to the linear in derivatives Lagrangian,
\begin{equation}\label{DirLarg}
  \mathcal{L}_\mathrm{D} = \bar{\psi} ( \mathrm{i} \gamma^\mu \partial_\mu - m ) \psi,
\end{equation}
where $\gamma^\mu = (\gamma^0,\bm{\gamma})$ are the Dirac matrices and $\bar{\psi} = \psi^\dag \gamma^0$. Based on Eq.~\eqref{DirLarg}, it is convenient to rewrite the Dirac equation in the form,
\begin{equation}\label{Direq}
  \dot\psi + (\bm{\alpha}\nabla) \psi + \mathrm{i}m\beta \psi = 0,
  \quad
  \dot\psi^* + (\bm{\alpha}^*\nabla) \psi^* - \mathrm{i}m\beta \psi^* = 0,
\end{equation}
where $\bm{\alpha} = \beta \bm{\gamma}$ and $\beta = \gamma^0$. We mention the following properties of the Dirac matrices: $\bm{\alpha}^\mathrm{T} = \bm{\alpha}^*$ and $\beta^\mathrm{T} = \beta$. The canonical momentum, conjugate to $\psi$, is $\pi = \tfrac{\partial \mathcal{L}_\mathrm{D}}{\partial \dot{\psi}} = \mathrm{i}\psi^*$.

Let us consider the Hamiltonian,
\begin{equation}\label{HamD}
  H_\mathrm{D} = \int \mathrm{d}^3 x
  \left\{
    (\psi^{*{}})^\mathrm{T} (\bm{\alpha}\nabla) \pi^{*{}}
    - \pi^\mathrm{T} (\bm{\alpha}\nabla) \psi    
    + \mathrm{i} m
    \left[
      (\psi^{*{}})^\mathrm{T} \beta \pi^*
      - (\pi)^\mathrm{T} \beta \psi
    \right]
  \right\},
\end{equation}
where the variables $\psi$, $\pi$, $\psi^*$, and $\pi^*$ are commuting $c$-numbers rather than Grassmann variables. The canonical equations based on Eq.~\eqref{HamD}, have the form,
\begin{align}
  \label{psiclass}
  \dot{\psi}  = & \frac{\delta H_\mathrm{D}}{\delta \pi} =
  -(\bm{\alpha}\nabla)\psi - \mathrm{i}m \beta \psi,
  \quad
  \dot{\psi}^*  = \frac{\delta H_\mathrm{D}}{\delta \pi^*} =
  - (\bm{\alpha}^\mathrm{T}\nabla)\psi^* + \mathrm{i}m \beta \psi^*,
  \\
  \label{piDclass}
  \dot{\pi}  = & - \frac{\delta H_\mathrm{D}}{\delta \psi} =
  - (\bm{\alpha}^\mathrm{T}\nabla)\pi + \mathrm{i}m \beta \pi,
  \quad
  \dot{\pi}^*  = - \frac{\delta H_\mathrm{D}}{\delta \psi^*} =
  - (\bm{\alpha}\nabla)\pi^* - \mathrm{i}m \beta \pi^*,
\end{align}
One can see that Eqs.~\eqref{psiclass} and~\eqref{piDclass} and equivalent to Eq.~\eqref{Direq}.

Now, using Eq.~\eqref{HamD}, we define the Lagrangian $\mathcal{L}_\mathrm{D}^{(\text{sym})} = \pi \dot{\psi} + \pi^* \dot{\psi}^* - \mathcal{H}$. Applying Eq.~\eqref{ELe} to $\mathcal{L}_\mathrm{D}^{(\text{sym})}$ with respect to $\psi$ and $\psi^*$, one reproduces Eq.~\eqref{piDclass}. One can rewrite $\mathcal{L}_\mathrm{D}^{(\text{sym})}$ in the form,
\begin{align}\label{DirLargsym}
  \mathcal{L}_\mathrm{D}^{(\text{sym})} = & \mathrm{i} (\psi^*)^\mathrm{T} \dot{\psi} - \mathrm{i}\psi^\mathrm{T} \dot{\psi}^*
  + \mathrm{i} (\psi^*)^\mathrm{T} (\bm{\alpha}\nabla) \psi
  - \mathrm{i} \psi^\mathrm{T} (\bm{\alpha}^\mathrm{T}\nabla) \psi^*
  -2 m (\psi^*)^\mathrm{T} \beta \psi
  \notag
  \\
  & =
  \mathrm{i} \bar{\psi} \gamma^\mu (\partial_\mu \psi) - \mathrm{i} (\partial_\mu \bar{\psi} ) \gamma^\mu \psi - 2m \bar{\psi} \psi.
\end{align}
The Lagrangian $\mathcal{L}_\mathrm{D}^{(\text{sym})}$ in Eq.~\eqref{DirLargsym} was shown in Ref.~\cite{BerLifPit82} to be equivalent to $\mathcal{L}_\mathrm{D}$ in Eq.~\eqref{DirLarg}.

\section{Quantization of a Dirac field\label{sec:QUANTDIR}}

In this section, we carry out the quantization of a massive Dirac field studied in Sec.~\ref{sec:DIRCLASS}.

According to Eqs.~\eqref{psiclass} and~\eqref{piDclass},  the variables $\psi$ and $\pi^*$, as well as $\psi^*$ and $\pi$, obey the identical wave equations. That is why we use the following form in representing them as plane waves solutions:
\begin{align}\label{eq:planewaves}
  \psi(x) & =\frac{1}{\sqrt{2}}\int\frac{\mathrm{d}^{3}p}{(2\pi)^{3/2}}
  \left[
    a_{s}(\mathbf{p})u_{s}(\mathbf{p})e^{-\mathrm{i}px}+b_{s}^{\dagger}(\mathbf{p})v_{s}(\mathbf{p})e^{\mathrm{i}px}
  \right],
  \notag
  \\
  \psi^{*}(x) & =\frac{1}{\sqrt{2}}\int\frac{\mathrm{d}^{3}p}{(2\pi)^{3/2}}
  \left[
    a_{s}^{\dagger}(\mathbf{p})u_{s}^{*}(\mathbf{p})e^{\mathrm{i}px}+b_{s}(\mathbf{p})v_{s}^{*}(\mathbf{p})e^{-\mathrm{i}px}
  \right],
  \notag
  \\
  \pi(x) & =\frac{1}{\sqrt{2}}\int\frac{\mathrm{d}^{3}p}{(2\pi)^{3/2}}
  \left[
    c_{s}^{\dagger}(\mathbf{p})u_{s}^{*}(\mathbf{p})e^{\mathrm{i}px}+d_{s}(\mathbf{p})v_{s}^{*}(\mathbf{p})e^{-\mathrm{i}px}
  \right],
  \notag
  \\
  \pi^{*}(x) & =\frac{1}{\sqrt{2}}\int\frac{\mathrm{d}^{3}p}{(2\pi)^{3/2}}
  \left[
    c_{s}(\mathbf{p})u_{s}(\mathbf{p})e^{-\mathrm{i}px}+d_{s}^{\dagger}(\mathbf{p})v_{s}(\mathbf{p})e^{\mathrm{i}px}
  \right],
\end{align}
where $a_{s}(\mathbf{p})$, $b_{s}(\mathbf{p})$, $c_{s}(\mathbf{p})$, $d_{s}(\mathbf{p})$, and the conjugated quantities are the annihilation and creation operators since we consider quantized fields now. The basis bispinors in Eq.~\eqref{eq:planewaves} are 
\begin{equation}\label{eq:basisspin}
  u_{s}(\mathbf{p})=\sqrt{\frac{E+m}{2E}}
  \left(
    \begin{array}{c}
      w_{s}(\mathbf{p}) \\
      \frac{sp}{E+m}w_{s}(\mathbf{p})
    \end{array}
  \right),
  \quad
  v_{s}(\mathbf{p})=\sqrt{\frac{E+m}{2E}}
  \left(
    \begin{array}{c}
      \frac{sp}{E+m}w_{s}(\mathbf{p}) \\
      w_{s}(\mathbf{p})
    \end{array}
  \right),
\end{equation}
where $w_{s}(\mathbf{p})$, $s=\pm{}$, are given in Eq.~\eqref{eq:helamp}.

The energy of the Dirac field is computed by the substitution of the decomposition in Eq.~\eqref{eq:planewaves} to Eq.~\eqref{HamD}. Then, we use the following relations:
\begin{align}
  u_{s}^{\dagger}(\mathbf{p})(\bm{\alpha}\mathbf{p})u_{s'}(\mathbf{p})= &
  v_{s}^{\dagger}(\mathbf{p})(\bm{\alpha}\mathbf{p})v_{s'}(\mathbf{p})=
  \delta_{ss'}\frac{p^{2}}{E},
  \notag
  \\
  v_{s}^{\dagger}(\mathbf{p})(\bm{\alpha}\mathbf{p})u_{s'}(-\mathbf{p})= &
  u_{s}^{\dagger}(\mathbf{p})(\bm{\alpha}\mathbf{p})v_{s'}(-\mathbf{p})=\mathrm{i}s\delta_{s,-s'}\frac{pm}{E},
  \notag
  \\
  u_{s}^{\dagger}(\mathbf{p})\beta u_{s'}(\mathbf{p})= & -v_{s}^{\dagger}(\mathbf{p})\beta v_{s'}(\mathbf{p})=\delta_{ss'}\frac{m}{E},
  \notag
  \\
  v_{s}^{\dagger}(\mathbf{p})\beta u_{s'}(-\mathbf{p})= & -u_{s}^{\dagger}(\mathbf{p})\beta v_{s'}(-\mathbf{p})=\mathrm{i}s\delta_{s,-s'}\frac{p}{E}
\end{align}
where we take into account Eq.~\eqref{eq:helampprop}. Eventually, we rewrite the energy in the form,
\begin{equation}\label{eq:Eindef}
  \mathcal{E}=\frac{\mathrm{i}}{2}\int\mathrm{d}^{3}pE
  \left[
    a_{s}^{\dagger}(\mathbf{p})c_{s}(\mathbf{p})-c_{s}^{\dagger}(\mathbf{p})a_{s}(\mathbf{p})
    +d_{s}(\mathbf{p})b_{s}^{\dagger}(\mathbf{p})-b_{s}(\mathbf{p})d_{s}^{\dagger}(\mathbf{p})
  \right].
\end{equation}
In Eqs.~\eqref{eq:basisspin}-\eqref{eq:Eindef}, $E = \sqrt{p^2 + m^2}$ in the energy of a free particle.

The momentum of the Dirac field has the form (see, e.g., Ref.~\cite{Dvo12a}),
\begin{equation}\label{eq:momDirgen}
  \bm{P} =\int\mathrm{d}^{3}x\left[(\psi^{*})^{\mathrm{T}}\nabla\pi^{*}-\pi{}^{\mathrm{T}}\nabla\psi\right].
\end{equation}
Analogously to Eq.~\eqref{eq:Eindef}, one can rewrite Eq.~\eqref{eq:momDirgen} as
\begin{equation}\label{eq:Pindef}
  \bm{P} =\frac{\mathrm{i}}{2}\int\mathrm{d}^{3}p\,\mathbf{p}
  \left[
    a_{s}^{\dagger}(\mathbf{p})c_{s}(\mathbf{p})-c_{s}^{\dagger}(\mathbf{p})a_{s}(\mathbf{p})
    +d_{s}(\mathbf{p})b_{s}^{\dagger}(\mathbf{p})-b_{s}(\mathbf{p})d_{s}^{\dagger}(\mathbf{p})
  \right],
\end{equation}
if we use the identities, $u_{s}^{\dagger}(\mathbf{p})u_{s'}(\mathbf{p})=v_{s}^{\dagger}(\mathbf{p})v_{s'}(\mathbf{p})=\delta_{ss'}$ and $v_{s}^{\dagger}(\mathbf{p})u_{s'}(-\mathbf{p})=u_{s}^{\dagger}(\mathbf{p})v_{s'}(-\mathbf{p})=0$,
which result from Eq.~\eqref{eq:basisspin}.

Up to now, the operators in Eqs.~\eqref{eq:Eindef} and~\eqref{eq:Pindef} are arbitrary. We impose the following constraint on them:
\begin{equation}\label{eq:relation}
  c_{s}(\mathbf{p})=-\mathrm{i}a_{s}(\mathbf{p}),
  \quad
  d_{s}(\mathbf{p})=\mathrm{i}b_{s}(\mathbf{p}),
\end{equation}
Note that Eq.~\eqref{eq:relation} is consistent with the fact that $\pi = \mathrm{i}\psi^*$. We also take that the independent operators obey the anticommutation relations,
\begin{equation}\label{eq:anticomm}
  \{ a_{s}(\mathbf{p}),a_{s'}^\dag(\mathbf{q}) \} = \delta_{ss'} \delta (\mathbf{p} - \mathbf{q} ),
  \quad
  \{ b_{s}(\mathbf{p}),b_{s'}^\dag(\mathbf{q}) \} = \delta_{ss'} \delta (\mathbf{p} - \mathbf{q} ),
\end{equation}
with the rest of the anticommutators being equal to zero.

Finally, using Eqs.~\eqref{eq:relation} and~\eqref{eq:anticomm}, we transform Eqs.~\eqref{eq:Eindef} and~\eqref{eq:Pindef} to the forms,
\begin{equation}\label{eq:Efin}
  \mathcal{E}=\int\mathrm{d}^{3}pE
  \left[
    a_{s}^{\dagger}(\mathbf{p})a_{s}(\mathbf{p})+b_{s}^{\dagger}(\mathbf{p})b_{s}(\mathbf{p})
  \right]
  +\dotsc,
\end{equation}
and
\begin{equation}\label{eq:Pfin}
  \bm{P} =\int\mathrm{d}^{3}p\,\mathbf{p}
  \left[
    a_{s}^{\dagger}(\mathbf{p})a_{s}(\mathbf{p})+b_{s}^{\dagger}(\mathbf{p})b_{s}(\mathbf{p})
  \right]
  +\dotsc.
\end{equation}
In Eqs.~\eqref{eq:Efin} and~\eqref{eq:Pfin}, we omit divergent vacuum terms. One can see in Eqs.~\eqref{eq:Efin} and~\eqref{eq:Pfin} that the energy and momentum of the Dirac field is the sum of contributions from the independent oscillators corresponding to particles and antiparticles. 

\section{Conclusion\label{sec:CONCL}}

In the present work, we have analyzed some features in the quantization of fermions in vacuum and in external fields. First, in Sec.~\ref{sec:QUANT}, we have constructed the quantum theory of a massive Majorana neutrino interacting with background matter. Starting from the Lagrangian for a classical Weyl field, we have derived the expressions for the total energy and the momentum of the field after it was secondly quantized. Then, in Sec.~\ref{sec:PROP}, we have derived the propagators of a massive Weyl neutrino in background matter. These results are used in Ref.~\cite{Dvo25} where neutrino flavor oscillations in matter are studied.

In Secs.~\ref{sec:QUANT} and~\ref{sec:PROP}, we have implicitly assumed that a Weyl field is represented in terms of anticommuting variables on the classical level. The alternative approach, where classical fermionic fields are given in terms of commuting $c$-numbers wavefunctions, was developed in Refs.~\cite{Dvo12a,Dvo12b}. This formalism is based on the Hamilton dynamics rather than on the Lagrange formalism. We clarify some of the issues of this approach in Sec.~\ref{sec:CLMAJ} applying it for massive Majorana neutrinos in background matter represented in terms of Weyl spinors. Some of the inexactitudes made in Ref.~\cite{Dvo12a} have been corrected in Sec.~\ref{sec:CLMAJ}.

Finally, Sec.~\ref{sec:DIRCLASS}, we apply the method developed in Ref.~\cite{Dvo12a} for the description of a massive Dirac fermion in vacuum. We have derived the Hamiltonian for classical $c$-number canonical variables which generates the known Dirac equation. This system has been quantized in Sec.~\ref{sec:QUANTDIR}, where we have obtained the correct form of the total energy and the momentum of a Dirac field.

In summary, we mention the commonly recognized treatment of classical fermionic fields is undoubtedly based on anticommuting Grassmann variables (see, e.g., Ref.~\cite{GitTyu90}). However, the results, obtained in the present work, provide an alternative point of view on this issue. The construction of the consistent theory of fermionic fields, represented in frames of our formalism, interacting with other particles, requires a separate special attention.

\end{document}